# Observation of pressure-induced Weyl state and superconductivity in a chirality-neutral Weyl semimetal candidate $SrSi_2$

M.-Y. Yao[1,*,6], J. Noky[1,6], Q.-G. Mu[1,2,3,6], K. Manna[1,4], N. Kumar[1], V. N. Strocov[5], C. Shekhar[1], S. Medvedev[1], Y. Sun[1], C. Felser[1,†]

[1]*Max Planck Institute for Chemical Physics of Solids, 01187 Dresden, Germany.*

[2]*Information Materials and Intelligent Sensing Laboratory of Anhui Province, Institutes of Physical Science and Information Technology, Anhui University, Hefei 230601, China*

[3]*Key Laboratory of Structure and Functional Regulation of Hybrid Materials of Ministry of Education, Anhui University, Hefei 230601, China*

[4]*Indian Institute of Technology- Delhi, Hauz Khas, New Delhi 110 016, India.*

[5]*Swiss Light Source, Paul Scherrer Institut, CH-5232 Villigen PSI, Switzerland.*

* E-mail: Mengyu.Yao@cpfs.mpg.de

† E-mail: Claudia.Felser@cpfs.mpg.de

[6] These authors contributed equally.

**Quasi-particle excitations in solids described by the Weyl equation have attracted significant attention in recent years. Thus far, a wide range of solids that have been experimentally realized as Weyl semimetals (WSMs) lack either mirror or inversion symmetry. For the first time, in the absence of both mirror and inversion symmetry, $SrSi_2$ has been predicted as a robust WSM by recent theoretical works. Herein, supported by first-principles calculations, we present systematic angle-resolved photoemission studies of undoped $SrSi_2$ and Ca-doped $SrSi_2$ single crystals. Our results show no evidence of the predicted Weyl fermions at the $k_z = 0$ plane or the Fermi arcs on the (001) surface. With external pressure, the electronic band structure evolved and induced Weyl fermions in this compound, as revealed by first-principle calculations combined with electrical transport property measurements. Moreover, a superconducting transition was observed at pressures above 20 GPa. Our investigations indicate that the $SrSi_2$ system is a good platform for studying topological transitions and correlations with superconductivity.**

There has been a surge of interest in gapless topological phases that arise in multiband fermionic systems, where the band-crossing points have nonzero Chern numbers [1-18]. This interest led to the discovery of quasiparticle excitations of Weyl fermions, which can be described using the same equations as their counterparts in high-energy physics [19]. By breaking either the time-reversal or inversion symmetry, Weyl fermions are realized in solids as two linearly dispersed crossing bands. The topology of Weyl nodes is characterized by a Chern number $C$, which is related to the Berry curvature. Another important hallmark of Weyl semimetals (WSMs) is the topologically protected Fermi arcs. In contrast to the bulk bands related to Weyl nodes, Fermi arcs are surface states that terminate at the projections of Weyl nodes with opposite chiral charges. Of note, the chiral charge of a Weyl node can be determined by counting connected Fermi arcs [17-18]. In recent years, WSMs have been experimentally identified in a wide range of materials, such as TaAs [2,4], $WP_2$ [10], $Co_3Sn_2S_2$ [11], and $Co_2MnGa$ [12]. Although quasiparticles within electronic band structures do not necessarily follow the Poincare symmetry pertaining to high-energy physics, they do adhere to crystal symmetry. Therefore, the stability of the WSM crystal structure is crucial for the formation of Weyl fermions. Moreover, they are important in device manufacturing and other applications.

Among the realized WSMs, Weyl fermions are mostly protected by either inversion or mirror symmetry. The inversion-asymmetric chiral compound strontium disilicide, $SrSi_2$, was predicted to be a robust WSM and the first that lacks both mirror and inversion symmetries [20-21]. $SrSi_2$ crystallizes in a simple cubic lattice system with a lattice constant of a = 6.563 Å and a space group of $P4_132$. As shown in Fig. 1(a), the crystal lacks both inversion and mirror symmetry. First-principles band structure calculations without spin-orbit coupling (SOC) in the generalized-gradient approximation show one electron band and one hole band along the Γ-X direction in the Brillouin zone (BZ). These two bands have an overlap of approximately 0.25 eV with two crossing points, as the calculated band structure shows in Fig. 1(c), which create a pair of Weyl fermions of opposite charges, as indicated by the dots in two colors shown in Fig. 1(b). In the presence of SOC, the linear band dispersion near the Weyl nodes becomes quadratic. As a result, the Weyl nodes transform into double-chiral states, which carry a higher chiral charge of ±2. In addition to the Weyl nodes, these first-principles calculations also predict long Fermi arcs across the entire BZ on the (001) surface. Furthermore, another theoretical work suggests that, with Ca doping, lattice shrinking increases the overlap between the hole and electron bands. As a result, the topological strength of $SrSi_2$ is expected to increase [21]. However, to the best of our knowledge, because of the difficulty of sample growth [22], no experimental research has been conducted on $SrSi_2$ systems. Direct evidence of the existence or absence of Weyl nodes remains lacking.

Herein, supported by first-principles band structure calculations, we present a comprehensive angle-resolved photoemission spectroscopy (ARPES) investigation of high-quality single-crystal $SrSi_2$ and Ca-doped $SrSi_2$. The ARPES experiments were carried out using vacuum-ultraviolet ARPES (VUV-ARPES) at the Bloch beam line at MAX IV with a DA30L analyzer and soft X-ray ARPES (SX-ARPES) at the ADRESS beamline at the Swiss Light Source [23] with a PHOIBOS-150 analyzer [24]. The data were collected using photon energies in the ultraviolet and soft X-ray regions, with overall energy resolutions of the order of 15 and 50−80 meV, respectively [25]. For the *ab initio* calculations, we employed density functional theory (DFT) as implemented in the Vienna *ab initio* simulation package (VASP) [26] using pseudopotentials and augmented plane waves as the basis set. The exchange-correlation potential was calculated using the hybrid HSE approximation [27]. To allow for a more detailed investigation through the BZ, maximally localized Wannier functions were created from the DFT wave functions using the program Wannier90 [28]. With the resulting tight-binding Hamiltonian, a dense search for possible Weyl crossings was performed. In a systematic study, we found no evidence of predicted Weyl nodes from ARPES at the $k_z = 0$ plane in the BZ for either undoped or Ca-doped $SrSi_2$ samples. This is in agreement with our HSE calculations. Moreover, we report the effects of high pressure on the electrical transport properties of $SrSi_2$ under external pressures of up to 65 GPa.

A semiconductor-metal transition occurred at approximately 12 GPa, which is comparable to our calculation using hybrid functional HSE results. Therefore, we suggest that $SrSi_2$ is a topological-trivial semiconductor at ambient pressure that becomes a WSM under increased pressure.

Samples with a typical size of approximately $8 \times 2 \times 2$ mm$^3$ were cleaved in situ at 15 K in a high-vacuum chamber with a base vacuum better than $5 \times 10^{-11}$ Torr. Core-level photoemission measurements of $SrSi_2$ were acquired with $h\nu$ = 200 eV, as shown in Fig. 2(a). The ARPES spectrum showed the characteristic core levels of Sr and Si, confirming the chemical composition of the samples. Using VUV-ARPES, we first performed a Fermi surface (FS) mapping on the (001) surface with $h\nu$ = 118 eV, as shown in Fig. 2(c). Four FS pockets were located along the Γ-X direction because of $C_4$ rotation symmetry. The out-of-plane FS was obtained with $h\nu$ in the range of 50−140 eV, as shown in Fig. 2(b). Owing to the Heisenberg uncertainty principle, the finite photoelectron escape depth results in a finite intrinsic resolution in the out-of-plane momentum $k_z$, with the $k_z$ broadening inversely proportional to the escape depth [24]. The photoelectrons emitted by ultraviolet photons have a relatively low escape depth. Therefore, the ARPES intensity was integrated over a relatively large $\Delta k_z$, and the out-of-plane FS contours were broadened along the $k_z$ direction. As a result, the FS pockets shown in the out-of-plane FS were ellipsoidal rather than the circular shape seen in the in-plane FS. However, both the in-plane and out-of-plane FS contours showed no signature of the predicted Fermi arcs of the surface states, albeit strongly depending on the terminations and surface chemical environment, in contrast to the bulk states. As the Fermi arcs are produced by Weyl nodes, the topological nature of $SrSi_2$ is questionable.

Because the Weyl nodes are band crossings of the bulk states, we then turned to the bulk states of $SrSi_2$ measured with SX-ARPES to achieve a higher bulk sensitivity and significantly improved $k_z$ resolution. The out-of-plane FS of $SrSi_2$, as shown in Fig. 3(a), was acquired within the $k_x$-$k_z$ plane with photon energies in the range of 580−820 eV. Similarly, for the out-of-plane FS obtained by VUV-ARPES, each BZ in Fig. 3(a) encloses four FS pockets along the high symmetry Γ-X direction. In Fig. 2(c), we present the in-plane FS map at $k_z = 0$ acquired with $h\nu = 704$ eV photons. Owing to the cubic crystal structure, the in-plane FS map is consistent with the out-of-plane FS map shown in Fig. 3(b), which also shows four FS pockets along the Γ-X direction. To determine the predicted Weyl nodes shown in Fig. 1(c), a high-resolution spectrum along the Γ-X direction was acquired in the $k_z = 0$ plane with $h\nu = 704$ eV, as shown in Fig. 3(c). Only one hole band was observed along this direction. The top of the hole band was located just above the Fermi energy ($E_F$), and it formed the FS pockets, which we saw in the in-plane and out-of-plane FSs. However, no electron band was observed to cross the hole band. To study the unoccupied states, we applied first-principles calculations using a hybrid functional HSE. As shown in Fig. 3(d), the calculated band structure along the Γ-X direction was in excellent agreement with the ARPES spectrum. In the unoccupied states, the energy maximum of the hole bands was located at approximately 50 meV above $E_F$, with an energy gap of approximately 80 meV. This gap size was experimentally confirmed and found to be of the same order. By analyzing the transport measurement within the range of 200 K to 270 K, we estimated the gap size to be approximately 23 meV at 0.4 GPa, as shown in Fig. 3(e). Therefore, the bottom of the electron bands was located above the hole bands, without crossing them.

In addition, to study the band structure evolution by the dopant, we performed SX-ARPES measurements on two Ca-doped $Sr_{1-x}Ca_xSi_2$ samples with x = 2% and 5%. As shown in Fig. 4(a) and (b), the ARPES spectra of the two doped samples exhibited identical band structures, which were also similar to those of the undoped sample (Fig. 3(c)). We show the momentum distribution curves (MDCs) taken at $E_F$ in Fig. 4(c) and (d). By measuring the distance between the MDC peaks, we found an average diameter of the hole FS pocket along the Γ-X direction of 0.09 Å$^{-1}$ for the 2% Ca-doped $SrSi_2$. This was slightly higher than the 0.073 Å$^{-1}$ measured for the 5% Ca-doped $SrSi_2$, which led to a larger FS pocket area and thus a higher metallicity of the less-doped sample. Doping-induced changes in the band structure depend on many factors, such as the lattice structure and chemical pressure. Compared with the undoped sample, the $Sr_{1-}$ $_xCa_xSi_2$ samples with x = 2% had a lattice constant reduction of 0.055%, which is negligible. Therefore, the lattice constant change is not the main factor in FS evolution by the Ca dopant.

To further study the band structure evolution with lattice distortion, we performed HSE calculations and electrical transport property measurements under high pressures. The effect of pressure on the electrical transport properties of $SrSi_2$ is illustrated in Fig. 5. At low pressures (below approximately 12 GPa), the resistivity curves demonstrated a non-monotonic temperature dependence (Fig. 5(a)), which is typical for narrow-gap semiconductors. At low temperatures (below 80–100 K), the concentration of charge carriers remained constant and phonon scattering resulted in increased resistivity with an increase in temperature, similar to metallic behavior. In contrast, at high temperatures, the intrinsic conductivity became activated and resistivity decreased with increasing temperature in accordance with semiconductor behavior. At a pressure of 0.6 GPa, the Hall-resistivity curve was linearly dependent on the magnetic field with a positive Hall coefficient (Fig. 5(b) and Fig. SM1(a)), indicating single-band hole-type charge carriers in $SrSi_2$, which is consistent with the data obtained at ambient pressure [22]. At pressures above 2.7 GPa, the Hall resistivity curves became nonlinear (Fig. 5(b) and Fig. SM1(b)) owing to the contribution of a second hole-pocket according to the two-band model fit (see details in Supplementary Materials), which is also evidenced by the electronic band structure calculations below. At pressures above 12 GPa, the temperature dependence of the resistivity exhibited a metallic character (Fig. 5(a)). This can be explained by the enhanced charge carrier density (Fig. SM1(e)) owing to the pressure-induced band overlap, as revealed by the electronic band structure calculations. Interestingly, as the pressure further increased, a superconducting transition with a critical temperature of $T_c^{10\%}$ approximately equal to 2.2 K was observed at approximately 21.8 GPa, and zero resistivity was achieved at approximately 27.4 GPa below 1.8 K (Fig. 5(a) and Fig. SM2(a)). The onset $T_c^{10\%}$ increased up to 5.0 K at 28.7 GPa and decreased slowly as the pressure further increased to 64.6 GPa at its highest (Fig. SM2(a) and (b)). However, the zero-resistivity temperature $T_c^{90\%}$ increased to 3.0

K at 34.1 GPa, and then became pressure insensitive (Fig. SM2(c)). Thus, the superconducting transition becomes sharper at higher pressures. Further studies on pressure-induced superconductivity are presented in the Supplementary Materials. Raman spectroscopy (Fig. SM3) demonstrated the absence of structural phase transitions in the entire studied pressure range, and the evolution of the crystal structure was reversible with respect to pressure. Thus, all the changes in the electrical transport properties described above are associated solely with the pressure-induced modification of the electronic band structure, as revealed by the theoretical calculations below.

Because no structural transition was observed in the Raman spectroscopy studies, we carried out HSE calculations with reduced lattice constants to simulate the external pressure. It can be seen that a smaller lattice constant results in a decrease in the calculated bandgap. At approximately 1.7% reduction, which corresponds to approximately 3.5 GPa, the gap vanished, and the system entered a metallic state, as shown in Fig. 5(c). At approximately 2.1%, corresponding to 4.4 GPa, the valence and conduction bands began to overlap and cross along the Γ-X direction. By theoretical analysis, we confirmed that these crossings were topologically non-trivial. Therefore, we suggest that the previously predicted Weyl nodes were recovered under pressure. The quantitative discrepancy between the theoretically predicted and experimentally observed values of the pressure of metallization might originate from the athermal nature of the electronic band calculations (that is, at $T = 0$ K, neglecting all entropic contributions) and well-known “overstabilization” of metallic states by density functional calculations [29].

Although no Weyl nodes were observed in the ARPES experiments below $E_F$, we could not rule out the existence of the predicted band crossings above $E_F$. In both scenarios, the FS pockets along the Γ-X direction enclosed either no Weyl nodes or two Weyl nodes with opposite chiral charges, leading to a net Chern number of $C = 0$. Furthermore, there was no evidence of the topological nature of $SrSi_2$ upon Ca doping. We suggest that HSE is a more suitable method for describing the exchange correlation for $SrSi_2$. Our first-principles calculations showed a gap, approximately 80 meV between the electron and hole bands, indicating that $SrSi_2$ is a topologically trivial semiconductor. Upon applying pressure, the evolution of the electrical transport properties was in good agreement with the calculations. When the lattice constant was reduced above 2.1%, we detected the topological phase transition by recovering the previously predicted Weyl nodes.

In conclusion, our study provides compelling evidence that $SrSi_2$ is a topologically trivial semiconductor with a band gap. While no Fermi arcs or Weyl nodes were observed in the ARPES experiments at ambient pressure, we find strong evidence for the existence of Weyl nodes under pressure using both experimental and theoretical analysis. Our results suggest that $SrSi_2$ can become a Weyl semimetal under pressure, and we observed pressure-induced superconductivity at high pressure.

Our findings challenge the previously predicted Weyl semimetallic nature of $SrSi_2$ and highlight the importance of considering pressure-induced changes in electronic properties in materials research. We believe that our study could have important implications for the development of new materials with interesting electronic properties, and we encourage further research in this field, including the use of ARPES with high-pressure sample holders as a tool for resolving similar problems.

**Acknowledgments**

This study was financially supported by an Advanced Grant from the European Research Council (No. 742068) 'TOPMAT', the European Union's Horizon 2020 research and innovation programme (No. 824123) 'SKYTOP', the European Union's Horizon 2020 research and innovation programme (No. 766566) 'ASPIN', the Deutsche Forschungsgemeinschaft (Project-ID 258499086) 'SFB 1143', the Deutsche Forschungsgemeinschaft (Project-ID FE 633/30-1) 'SPP Skyrmions', the DFG through the Würzburg-Dresden Cluster of Excellence on Complexity and Topology in Quantum Matter ct.qmat (EXC 2147, Project-ID 39085490), and the National Natural Science Foundation of China (No. 12204007). We acknowledge Balasubramanian Thiagarajan and Craig Polley for their help with ARPES experiments.

## Author contributions

The ARPES measurements were conducted by M.-Y. Y. and supported by V.N.S. First-principles calculations were performed by J.N. and Y.S. High-pressure transport measurements and Raman spectra were obtained by Q.-G.M. and S.M. The single crystals were grown by K.M. with the help of N.K. and C.S. All authors have discussed the results. The manuscript was written by M.-Y.Y. with feedback from all authors. The project was supervised by C.F.

## Competing financial interests

The authors declare no competing financial interests.

## References

[1] S.-Y. Xu, C. Liu, S. K. Kushwaha, R. Sankar, J. W. Krizan, I. Belopolski, M. Neupane, G. Bian, N. Alidoust, T.-R. Chang et al., *Science* **347**, 294 (2015).

[2] B. Q. Lv, S. Muff, T. Qian, Z. D. Song, S. M. Nie, N. Xu, P. Richard, C. E. Matt, N. C. Plumb, L. X. Zhao et al., *Phys. Rev. Lett.* **115**, 217601 (2015).

[3] N. Xu, H. M. Weng, B. Q. Lv, C. E. Matt, J. Park, F. Bisti, V. N. Strocov, D. Gawryluk, E. Pomjakushina, K. Conder et al., *Nat. Commun.* **7**, 11006 (2015).

[4] S.-Y. Xu, I. Belopolski, N. Alidoust, M. Neupane, G. Bian, C. Zhang, R. Sankar, G. Chang, Z. Yuan, C.-C. Lee et al., *Science* **349**, 613–617 (2015).

[5] B. Q. Lv, N. Xu, H. M. Weng, J. Z. Ma, P. Richard, X. C. Huang, L. X. Zhao, G. F. Chen, C. E. Matt, F. Bisti et al., *Nat. Phys.* **11**, 724–727 (2015).

[6] H. Weng, C. Fang, Z. Fang, B. A. Bernevig, and X. Dai, *Phys. Rev. X* **5**, 11029 (2015).

[7] B. Bradlyn, J. Cano, Z. Wang, M. G. Vergniory, C. Felser, R. J. Cava, and B. A.

Bernevig, *Science* **353**, 5037 (2016).

[8] N. Xu, G. Autès, C. E. Matt, B. Q. Lv, M. Y. Yao, F. Bisti, V. N. Strocov, D. Gawryluk, E. Pomjakushina, K. Conder et al., *Phys. Rev. Lett.* **118**, 106406 (2017).

[9] S.-Y. Xu, N. Alidoust, G. Chang, H. Lu, B. Singh, I. Belopolski, D. S. Sanchez, X. Zhang, G. Bian, H. Zheng et al., *Sci. Adv.* **3**, e1603266 (2017).

[10] M.-Y. Yao, N. Xu, Q. S. Wu, G. Autès, N. Kumar, V. N. Strocov, N. C. Plumb, M. Radovic, O. V. Yazyev, C. Felser et al., *Phys. Rev. Lett.* **122**, 176402 (2019).

[11] D. F. Liu, A. J. Liang, E. K. Liu, Q. N. Xu, Y. W. Li, C. Chen, D. Pei, W. J. Shi, S. K. Mo, P. Dudin et al., *Science* **365**, 1282–1285 (2019).

[12] I. Belopolski, K. Manna, D. S. Sanchez, G. Chang, B. Ernst, J. Yin, S. S. Zhang, T. Cochran, N. Shumiya, H. Zheng et al., *Science* **365**, 1278–1281 (2019).

[13] D. S. Sanchez, I. Belopolski, T. A. Cochran, X. Xu, J.-X. Yin, G. Chang, W. Xie, K. Manna, V. Süß, C.-Y. Huang et al., *Nature* **567**, 500–505 (2019).

[14] N. B. M. Schröter, D. Pei, M. G. Vergniory, Y. Sun, K. Manna, F. de Juan, J. A. Krieger, V. Süss, M. Schmidt, P. Dudin et al., *Nat. Phys.* **15**, 759–765 (2019).

[15] Z. Rao, H. Li, T. Zhang, S. Tian, C. Li, B. Fu, C. Tang, L. Wang, Z. Li, W. Fan, et al., *Nature* **567**, 496–499 (2019).

[16] D. Takane, Z. Wang, S. Souma, K. Nakayama, T. Nakamura, H. Oinuma, Y. Nakata, H. Iwasawa, C. Cacho, T. Kim et al., *Phys. Rev. Lett.* **122**, 076402 (2019).

[17] M. Yao, K. Manna, Q. Yang, A. Fedorov, V. Voroshnin, B. Valentin Schwarze, J. Hornung, S. Chattopadhyay, Z. Sun, S. N. Guin et al., *Nat. Commun.* **11**, 2033 (2020).

[18] N. B. M. Schröter, S. Stolz, K. Manna, F. De Juan, M. G. Vergniory, J. A. Krieger, D. Pei, T. Schmitt, P. Dudin, T. K. Kim et al., *Science* **369**, 179–183 (2020).

[19] H. Weyl, *Z. Phys.* **56**, 330–352 (1929).

[20] S.-M. Huang, S.-Y. Xu, I. Belopolski, C.-C. Lee, G. Chang, T.-R. Chang, B. Wang, N. Alidoust, G. Bian, M. Neupane et al., *Proc. Natl. Acad. Sci.* **113**, 1180–1185 (2016).

[21] B. Singh, G. Chang, T. R. Chang, S. M. Huang, C. Su, M. C. Lin, H. Lin, and A. Bansil, *Sci. Rep.* **8**, 1–9 (2018).

[22] K. Manna, N. Kumar, S. Chattopadhyay, J. Noky, M. Yao, J. Park, T. Forster, M. Uhlarz, B. V. Schwarze, J. Hornung et al., *Phys. Rev. B* **106**, L041113 (2022).

[23] V. N. Strocov, T. Schmitt, U. Flechsig, T. Schmidt, A. Imhof, Q. Chen, J. Raabe, R. Betemps, D. Zimoch, J. Krempasky et al., *J. Synchrotron Rad.* **17**, 631 (2010).

[24] V. N. Strocov, X. Wang, M. Shi, M. Kobayashi, J. Krempasky, C. Hess, T. Schmitt, and L. Patthey, *J. Synchrotron Radiat.* **21**, 32–44 (2014).

[25] V. N. Strocov, M. Shi, M. Kobayashi, C. Monney, X. Wang, J. Krempasky, T. Schmitt, L. Patthey, H. Berger, and P. Blaha, *Phys. Rev. Lett.* **109**, 086401 (2012).

[26] G. Kresse and J. Furthmüller, *Phys. Rev. B* **54**, 11169–11186 (1996).

[27] J. Heyd, G. E. Scuseria, and M. Ernzerhof, *J. Chem. Phys.* **118**, 8207 (2003).

[28] A. A. Mostofi, J. R. Yates, Y.-S. Lee, I. Souza, D. Vanderbilt, and N. Marzaria, *Comput. Phys. Commun.* **178**, 685–699 (2008).

[29] Y. Ma, A. R. Oganov, and C. W. Glass, *Phys. Rev. B* **76**, 064101 (2007).

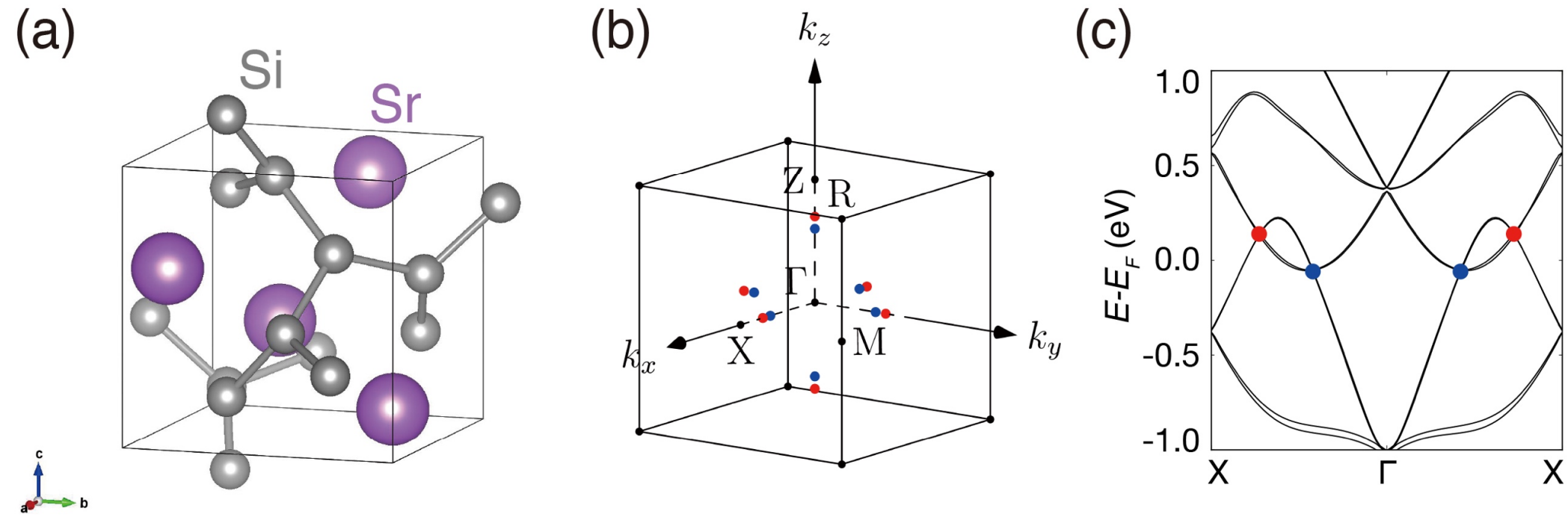

Fig. 1 (a) Crystal structure of $SrSi_2$. (b) Bulk BZ of $SrSi_2$ with high-symmetry points labeled. The blue and red dots represent the predicted Weyl nodes along Γ-X direction. (c) GGA Calculated band structure along the Γ-X direction. Due to a lack of mirror-symmetry, the Weyl nodes with opposite charges (blue and red dots) are located at different energies.

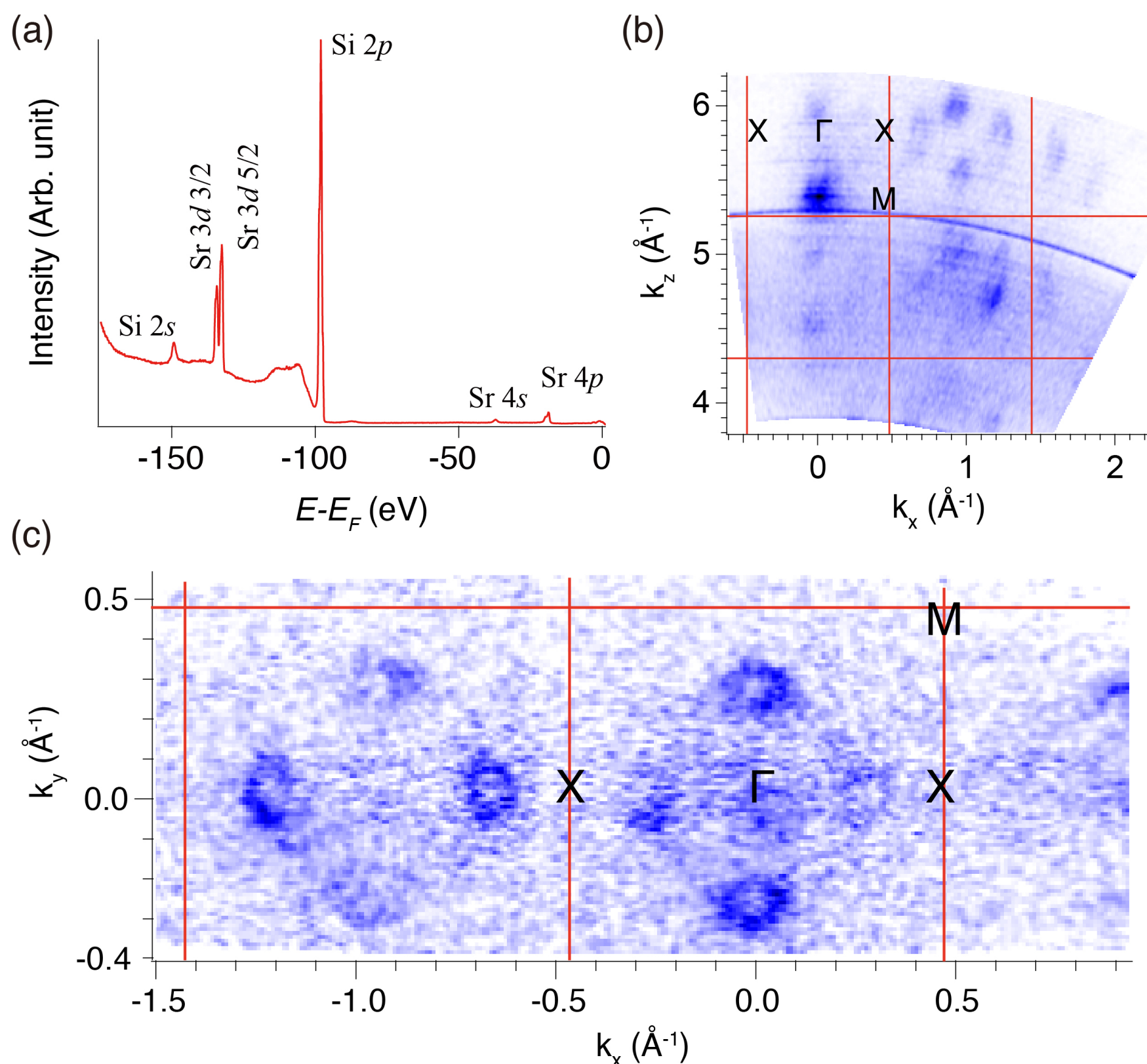

Fig. 2 (a) $SrSi_2$ core level spectrum, obtained with $h\nu$ = 200 eV. (b) The out-of-plane FS, taken with the photon energy in the range of 50−140 eV, with the BZ boundary marked (red lines). (c) The in-plane FS at $k_z$ = 0 plane, acquired with $h\nu$ = 118 eV.

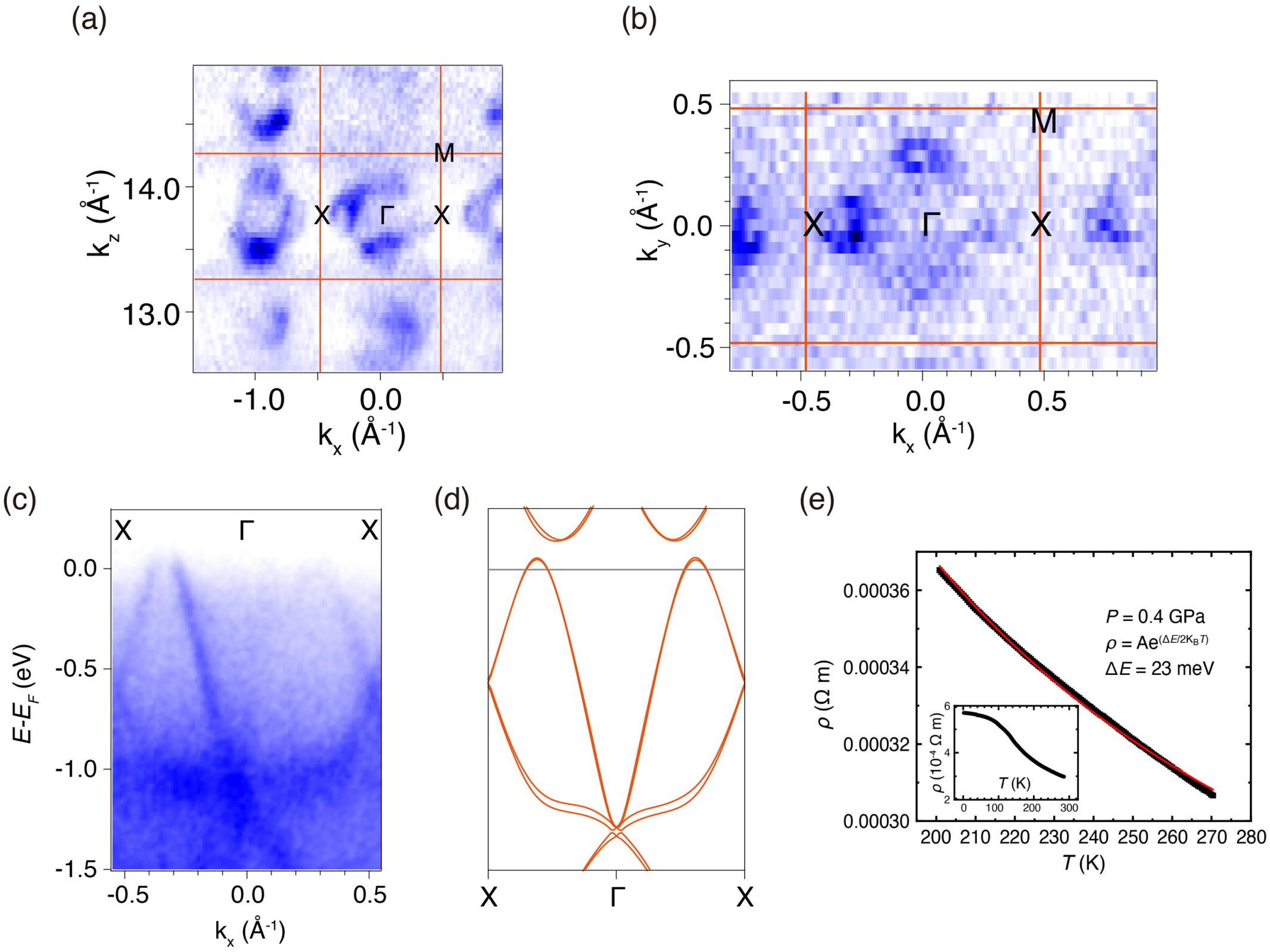

Fig. 3 (a) The out-of-plane and in-plane FS, acquired with photon energy in the range of 580−820 eV, with BZ boundary marked (red lines). (b) The in-plane FS at $k_z = 0$. (c-d) Photoemission intensity plot along the Γ-X direction and corresponding calculated band structure, respectively. (e) Resistivity measurement at pressure of 0.4 GPa demonstrated a semiconductor behavior at high temperature. A band gap of 23 meV was obtained by fitting data from 200 to 270 K with Arrhenius equation.

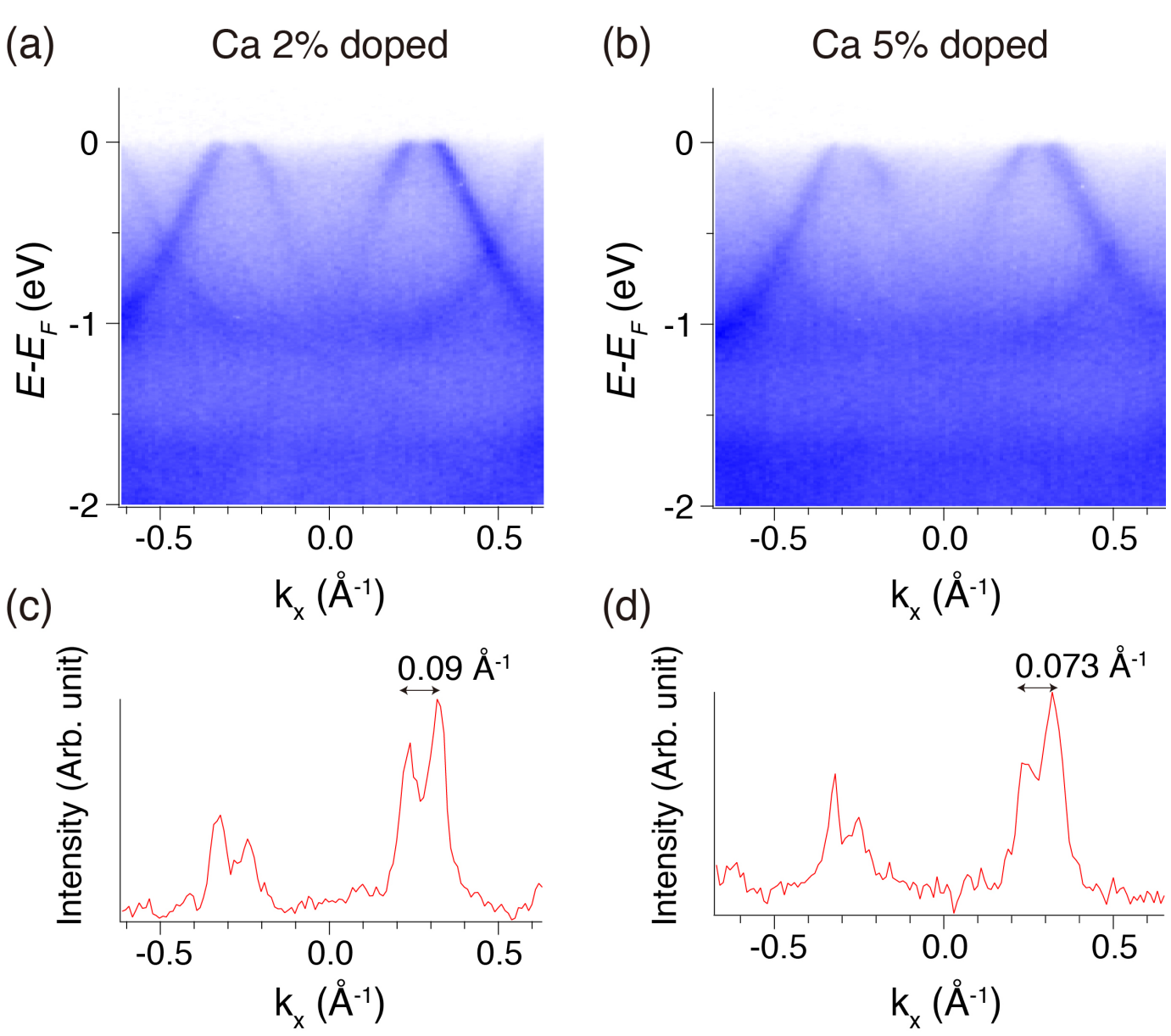

Fig. 4 (a) High resolution photoemission intensity plot of $Ca_xSr_{1-x}Si_2$ (x = 2%) along the Γ-X direction at the $k_z = 0$ plane, and (c) its MDC taken at $E_F$. (b,d) Same as (a,c) but for $Ca_xSr_{1-x}Si_2$ (x = 5%).

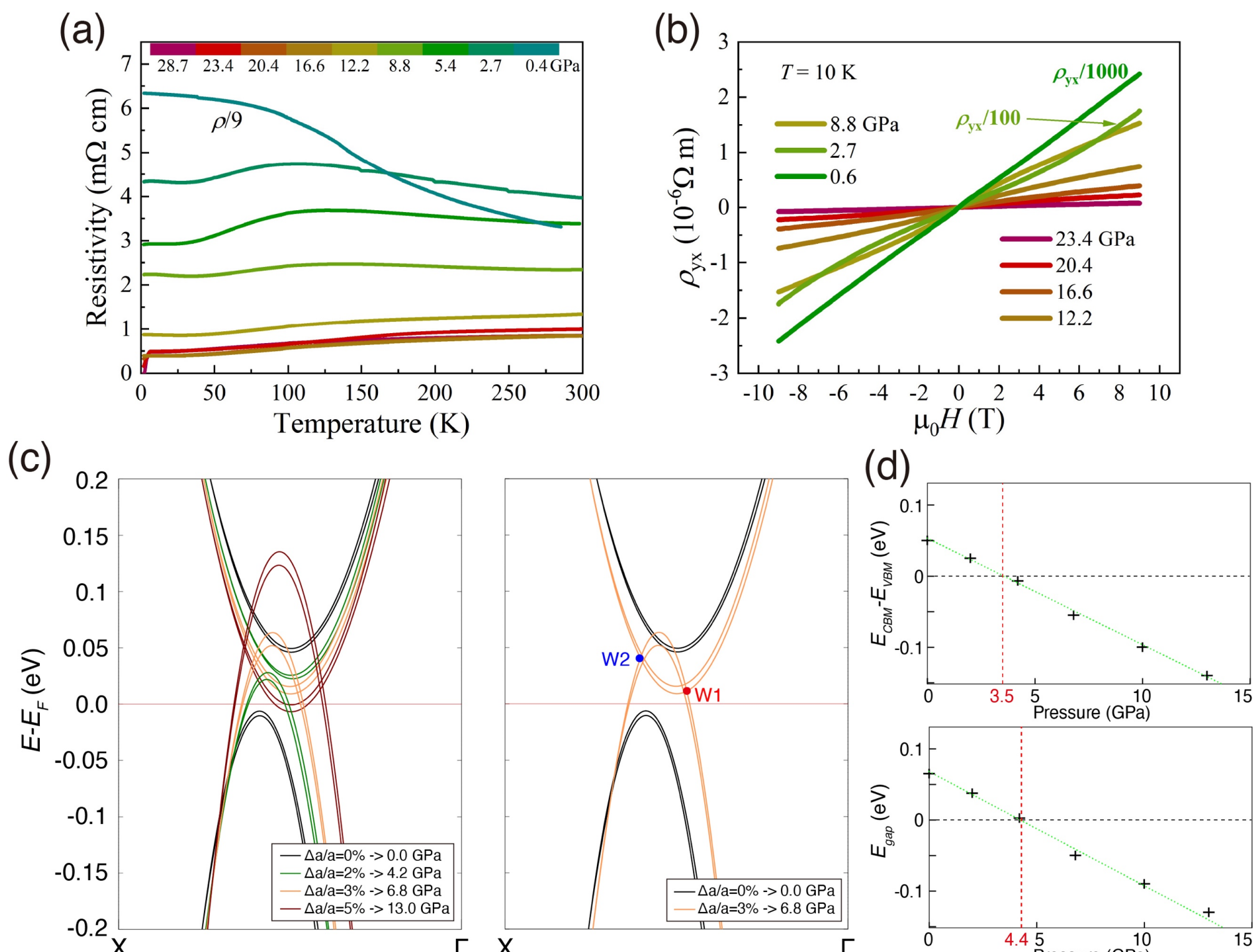

Fig. 5 (a) Temperature dependence of electrical resistivity with pressures up to 28.7 GPa. $SrSi_2$ shows a semiconductor-metal transition at approximately 12 GPa and undergoes a superconducting transition with pressure above 20.4 GPa. (b) Hall resistivity curves under selected pressures displaying a transition from linear to nonlinear field dependence, which suggests that charge carrier changes from single band hole type to multiband. (c) HSE calculations under varying pressure. (d) Calculated band gap changes with pressure.